\documentclass[graybox]{svmult}

\usepackage{makeidx}         
\usepackage{graphicx}        
\usepackage{multicol}        
\usepackage[bottom]{footmisc}

\usepackage{newtxtext}       %
\usepackage[varvw]{newtxmath}       

\usepackage[utf8]{inputenc}
\usepackage[english]{babel}
\usepackage{amsmath,amsthm,amsfonts}
\usepackage{multirow}
\usepackage{tikz}
\usepackage{verbatim}
\usepackage{todonotes}
\usepackage{footmisc}
\usepackage{tocloft} 
\usepackage{isotope}
\usepackage{braket}
\usepackage{siunitx}
\DeclareSIUnit\torr{Torr} 
\DeclareSIUnit\gauss{G}

\usepackage[T1]{fontenc}

\usepackage{hyperref}
\hypersetup{
    colorlinks=true,
    linkcolor=blue,
    filecolor=blue,      
    urlcolor=cyan,
    citecolor=blue,
    pdftitle={Quantum computing with atomic qubit arrays: confronting the cost of connectivity},
    pdfpagemode=FullScreen,
    }
\usepackage[hyperpageref]{backref}

\makeindex             

\newcommand \be{\begin{equation}}
\newcommand \ee{\end{equation}}
\newcommand \bea{\begin{eqnarray}}
\newcommand \eea{\end{eqnarray}}
\newcommand \bse{\begin{subequations}}
\newcommand \ese{\end{subequations}}

\newcommand \mcF{{\mathcal F}}

\begin{document}

\title*{Quantum computing with atomic qubit arrays: confronting the cost of connectivity}
\titlerunning{Quantum computing with atomic qubit arrays}
\author{M. Saffman\orcidID{0000-0001-6398-2097}}
\institute{Department of Physics, University of Wisconsin-Madison, Madison, WI 53706, USA
\and 
Infleqtion, Madison, WI, 53703 USA, \email{msaffman@wisc.edu}}
\maketitle

\abstract{These notes present a review of the status of quantum computing with arrays of neutral atom qubits, an approach which has demonstrated remarkable progress in the last few years. Scaling digital quantum computing to qubit counts and control fidelities that will enable solving outstanding scientific questions, and provide commercial value, is an outstanding challenge, not least because of the requirement of connecting and entangling distant qubits. Long-range Rydberg gates and physical motion outfit atomic qubit arrays with tools for establishing  connectivity. These tools operate on different timescales and with distinct levels of parallelization. We analyze several prototypical architectures from the perspective of achieving fast connectivity for circuits with large scale entanglement, as well as fast cycle times for measurement based quantum error correcting codes. Extending Rydberg interactions to multiple atomic species has emerged as a promising route to achieving this latter requirement.     }


\section{Introduction}

The concept of a quantum computer is not new  \cite{Feynman1982,Feynman1986, Manin1980, Deutsch1985}, yet the timeline for wide availability of quantum computers that are useful for solving problems in science or commerce is uncertain.\footnote{ 
The material presented here is based on lectures given at the  International School of Physics ``Enrico Fermi" in Varenna, Italy in July, 2024. Given the availability of other review papers and the very rapid pace of development of quantum computing with neutral atoms I have chosen not to include all the background material presented in the lectures. Instead some new developments up until summer 2025 when these notes were finalized are presented. }  The uncertainty does not arise from questions about the  predictive power of quantum mechanics, which was established 100 years ago and has been verified in countless experiments. Rather, scientists and engineers who are striving to build quantum computers do not yet have a verified blueprint for how to proceed.  It should also be acknowledged  that objections to the completeness of quantum mechanics  \cite{Einstein1935},
and its seemingly absurd predictions about the physical world  \cite{Schrodinger1935,Schrodinger1935b,Schrodinger1936}
have persisted  \cite{tHooft2021}, despite the success of experiments which have confirmed the violation of Bell's inequality  \cite{Giustina2015,Shalm2015,Rosenfeld2017} as predicted by quantum theory  \cite{Bell1964}. One of 
the fundamental applications of large scale quantum computing may well be the creation of new tools for  studying the foundations of quantum theory.

There is no single reason why developing a quantum computer is so challenging. Even so, a few obstacles stand out. Quantum computational advantage  derives from the possibility of controlling quantum phenomena including superposition and entanglement. Preserving these quantum phenomena demands that the qubits are well isolated from the environment. On the other hand a computer is not just a science experiment but is a machine that can be programmed to perform desired tasks, which requires control. Achieving both isolation and control in a single system is difficult and has been tackled most successfully in just a handful  of physical platforms, including trapped ions, superconducting circuits, photonics, electrons in low-dimensional semiconductors, and neutral atoms  \cite{Bergou2021}. Apart from the dichotomy of isolation and control there is the challenge of connectivity. In conventional computers information is freely copied and moved across circuits on wires. In the quantum world unknown states cannot be copied  \cite{Wootters1982,Dieks1982}, and can only be moved to a different physical location if no trace  of the state is left behind. As a consequence performing logic with  qubits that may be physically far apart is 
not trivial and significantly adds to the challenge of scaling up quantum computers. This challenge is present in all of the aforementioned platforms. Neutral atoms provide two mechanisms for achieving connectivity: long-range Rydberg interactions  \cite{Lukin2001,Saffman2010}, and physical transport of qubits  \cite{Bluvstein2022}. These two mechanisms operate on different spatial and temporal scales which has consequences for how to optimize the connectivity architecture.

\begin{figure}[!t]
\center
\includegraphics[width=.7\columnwidth]{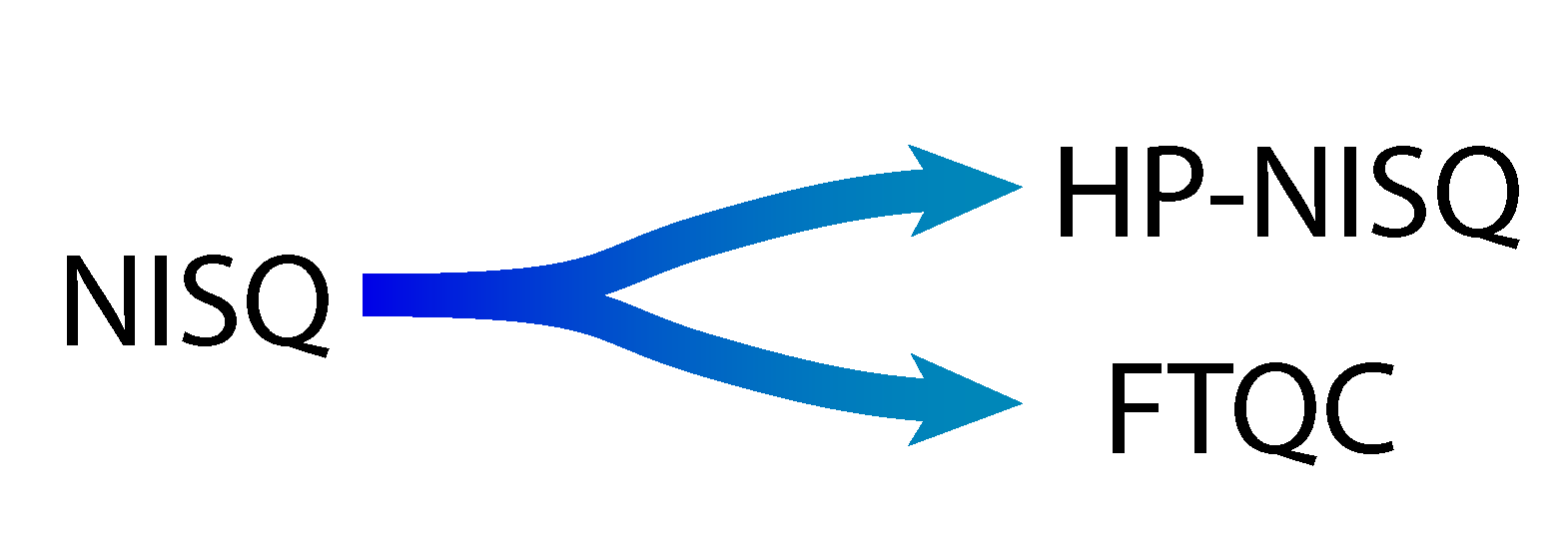}
\caption{\label{fig.hpnisq} A bifurcated scenario for the development of quantum computers beyond the current NISQ era. High performance NISQ (HP-NISQ) for beyond classical capabilities requires moderate increases in qubit count and a substantial increase in gate fidelity. Quantum error correction and Fault Tolerant Quantum Computing(FTQC) will require orders of magnitude more qubits and a moderate increase in gate fidelity.}
\end{figure}

Before proceeding to specifics of neutral atom qubits, it seems useful to delineate two distinct paths forward, and the implications for future progress. These paths are indicated in Fig. \ref{fig.hpnisq}. The current state of play has been termed the NISQ era for Noisy Intermediate Scale Quantum  \cite{Preskill2018}. Working quantum computers exist and have demonstrated impressive performance, even though they have a limited number of qubits and suffer from errors in control and measurements. Despite these limitations NISQ machines have performed  calculations 
based on random circuit sampling (RCS) that produce bitstrings with a statistical distribution that cannot be mimicked by conventional computers. These so-called ``Quantum Supremacy" experiments  \cite{Arute2019,YWU2021,Morvan2024,DeCross2025,DGao2025} demonstrate the capability of performing a particular calculation that is beyond the reach of conventional computers. but not yet a calculation that has wider utility. The possibility of using RCS not just to demonstrate the power of a quantum computer, but to solve useful problems, has not been decisively demonstrated, although there are promising opportunities involving applications of certified randomness  \cite{MLiu2025} and quantum fingerprinting for comparing remote databases  \cite{Gokhale2022}.  
Other demonstrations of NISQ hardware that seek to demonstrate  computational advantage for digital quantum circuits have targeted the transverse field Ising model  \cite{YKim2023,Haghshenas2025} which has broad applicability as a description of interactions and dynamics of many-body spin systems. Very recent results enabled by  high fidelity entangling gates simulated prethermalization in a regime beyond currently available classical techniques  \cite{Haghshenas2025}.

While  the progress in RCS and transverse field Ising simulations with quantum processors is remarkable, the progress has also inspired development of classical algorithms that successfully competed with early quantum demonstrations  \cite{FPan2022,Begusic2024}. This situation with a back and forth competition between  conventional computers and digital quantum hardware begs the question of the viability of  quantum computational advantage in the current NISQ era without quantum error correction and fault tolerance.  

To clarify this question we need to establish a connection between the fidelity of expectation values of observables provided by a quantum circuit and the classical cost of simulating the observable. The performance of state of the art NISQ processors is typically limited by the fidelity of the two-qubit gate used for entanglement. As discussed in  \cite{Kechedzhi2024} the expectation value of an operator $\hat O$ that is obtained from a noisy quantum state with density matrix $\rho$ can be expressed as ${\rm tr}(\rho\hat O)=\mcF_{\rm eff}\langle \hat O\rangle_{\rm ideal}$ where $\mcF_{\rm eff}$ is the effective fidelity of the observable and  $\langle \hat O\rangle_{\rm ideal}$ is the noise free expectation value. An estimate for the fidelity is 
\begin{equation}
 \mcF_{\rm eff}\sim e^{-\epsilon V_{\rm eff}}
 \label{eq.Feff}
 \end{equation}
 where $\epsilon$ is the two-qubit gate infidelity and $V_{\rm eff} $ is the circuit volume, which may be approximated by the number of two-qubit gates involved in calculating the expectation value. 
A tensor network simulation  \cite{Orus2014} of the expectation value on a classical computer has a computational cost that approximately scales as
\begin{equation}
C\sim 2^A
\label{eq.cost}
 \end{equation}
where $A$ is the effective area of a cut of the circuit volume, which is essentially a proxy for the maximum amount of entanglement generated.

\begin{figure}[!t]
\center
\includegraphics[width=.7\columnwidth]{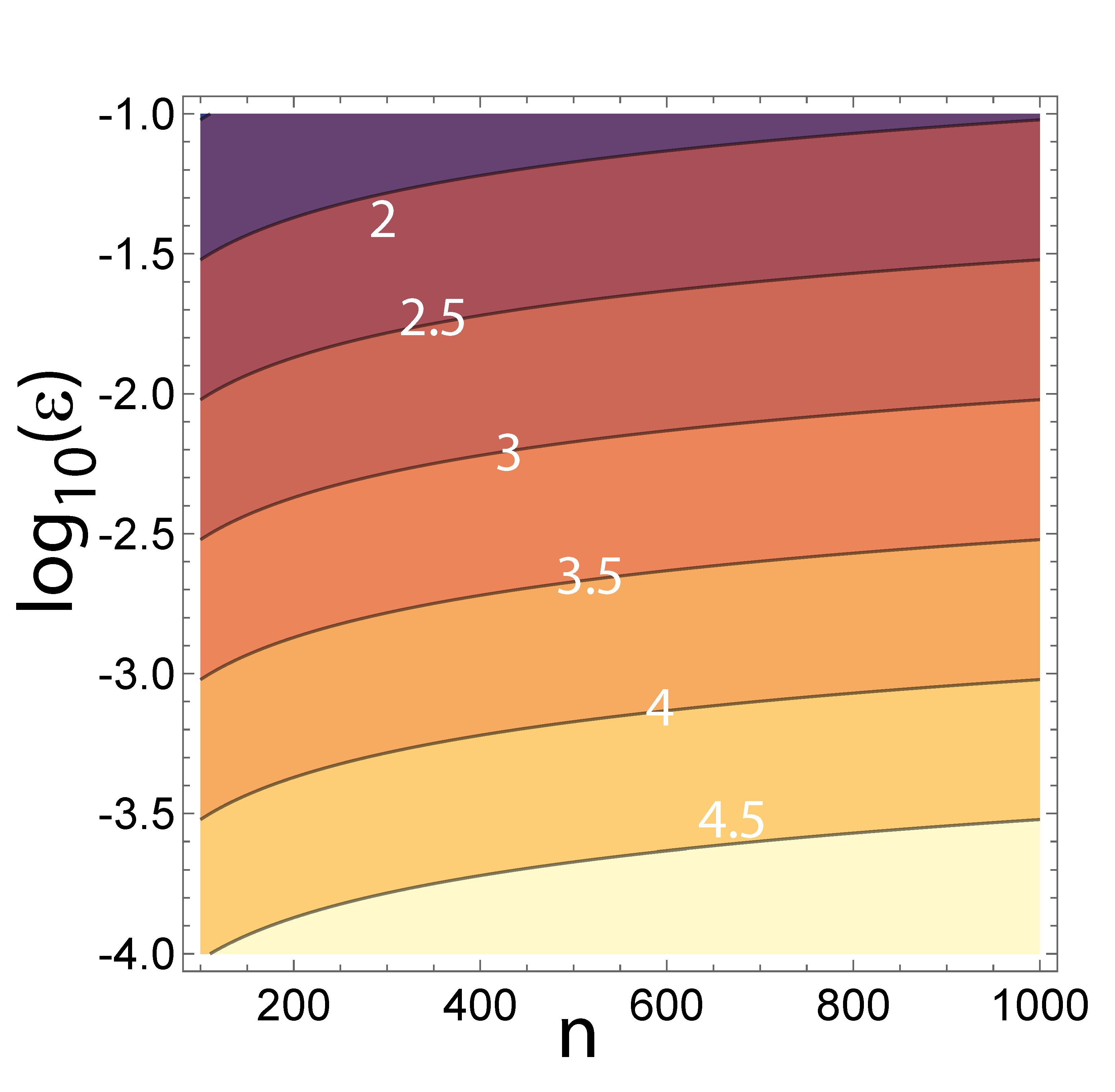}
\caption{\label{fig.cost} Contour plot of $\log_{10}(\log_{10}(C)),$ the double logarithm of the classical calculational cost as a function of number of qubits and gate infidelity.}
\end{figure}

The scaling of $A$ with the number of qubits and gate fidelity depends on a number of factors including the dimensionality and connectivity of the circuit, the type of entangling gate used, and the physical problem of interest. For RCS and short evolution times $A\sim \sqrt n t$ with $t$ the execution time (circuit depth) for a square array of $n$ qubits  \cite{Kechedzhi2024}. The scaling relations  (\ref{eq.Feff},\ref{eq.cost}) imply that a noisy circuit will output a fidelity that decreases exponentially with circuit volume, while the classical cost of simulating the result increases exponentially with the area of a circuit cut, which can be thought of as a proxy for the maximum amount of entanglement generated. Crucially the exponential scaling of the fidelity and the classical computational cost are different  \cite{Kechedzhi2024}. Putting $t\sim 1/\epsilon$ we have $C\sim 2^{\sqrt n / \epsilon}$ which has been visualized in Fig. \ref{fig.cost}. Experimental progress in the last few years has led to a rapid increase of the estimated equivalent computational cost of RCS performed on quantum hardware. For that reason recent papers  \cite{Morvan2024, DGao2025} choose to present $\log(\log(C))$ and we have made the same choice in Fig. \ref{fig.cost}. It can be seen that the contours of constant $C$ make a small angle with the abscissa  and have a large gradient aligned towards the ordinate. Thus the critical requirement for increasing the computational cost is the gate fidelity, not the number of qubits. This is consistent with other measures of computational power such as the quantum volume  \cite{Cross2019}. 

State of the art utility experiments are being performed with $n\sim 100$ and $\epsilon=1-\mcF\sim 0.001.$ It is apparent from Fig. \ref{fig.cost} that increasing the qubit count up to 1000 has only a moderate impact on the computational cost. On the other hand improving the gate fidelity by another factor of 10 will give a much more pronounced boost to the cost of a classical simulation. The implication is that the path to HP-NISQ suggested in Fig. \ref{fig.hpnisq} primarily requires increasing control fidelity. 
Realizing the full potential of quantum computers will undoubtedly require FTQC with orders of magnitude more qubits than HP-NISQ machines. While higher gate fidelity will help to reduce the requirements on qubit count, even at extremely high fidelities the code rate will at best be $1/5$  \cite{Laflamme1996} with fault tolerant universality needing additional qubits for magic state distillation  \cite{Bravyi2005}.

The rest of this contribution is organized as follows. In Sec. \ref{sec.scaling} we give a brief summary of the current state of the art for neutral atom, gate based quantum computers. Progress in scaling qubit count, and performing fast, high fidelity measurements is reviewed. As described above higher gate fidelity is crucial for realizing computational power without error correction. Some fundamental limits and prospects for increasing neutral atom gate fidelity are discussed in Sec. \ref{sec.fidelity}. Operation at the logical level with many more physical qubits brings the challenge of connectivity to the forefront. Several approaches are compared in Sec. \ref{sec.connectivity}. An additional requirement for logical qubit operations is mid-circuit measurements which are discussed with a focus on a two-species architecture in Sec. \ref{sec.midcircuit}. We conclude with a perspective on the present challenges and an  outlook for future developments in Sec. \ref{sec.outlook}.

\section{Status and scaling of neutral atom qubit arrays}
\label{sec.scaling}

In the last few years there has been a remarkable and  rapid increase in the rate of development of neutral atom quantum processors. Many new academic research groups have entered the field and there are more than five companies seeking to commercialize neutral atom quantum computing. Key performance metrics of state of the art systems include large 2D arrays with hundreds of qubits,  qubit coherence times of several seconds, gate operation times of a few $\mu\rm s$ to sub-$\mu\rm s$ for one- and two-qubit gates, single shot measurement and state preparation and measurement (SPAM) fidelity above 99\%. Gate fidelities have reached $\mcF>0.9999$ for global rotations driven by microwaves  \cite{CSheng2018,Nikolov2023}, $\mcF > 0.999$ for qubit specific single qubit gates  \cite{Radnaev2025,Chinnarasu2025}, and $\mcF > 0.99$ for two-qubit gates. Using long-range Rydberg interactions or atom motion, connectivity can extend beyond nearest neighbor as will be discussed in detail in Sec. \ref{sec.connectivity}.

With these advances neutral atom qubit arrays are now positioned as a leading modality in the race to realize useful quantum computing. Arrays with hundreds of qubits have been used for analog simulation of interacting spin models and solution of optimization problems  \cite{Scholl2021,Ebadi2021,Ebadi2022}. Gate model quantum circuits have demonstrated basic algorithms at the physical qubit level  \cite{Graham2022,Chinnarasu2025} as well as circuits with logical qubits  \cite{Bluvstein2022, Bluvstein2024, Bedalov2024, Reichardt2024}, circuit based preparation of entangled states for metrology  \cite{ACao2024,Finkelstein2024}, and first demonstrations of magic state distillation  \cite{Rodriguez2025}.  
Capabilities that are being actively improved upon include qubit counts, fast and low loss qubit measurements, and gate fidelities.

In terms of qubit counts there have been several demonstrations of atom arrays with more than 1000 trapping sites  \cite{Ebadi2021, Huft2022, Pichard2024, Norcia2024, Gyger2024} with the latest results achieving 2D arrays with 6100 atoms loaded on average into 12000 sites  \cite{Manetsch2025}, and 2000 atoms deterministically rearranged into desired patterns with almost no defects  \cite{RLin2025}. 
The Talbot effect  \cite{Talbot1836}  leads to axial repetition of transversally periodic structures and can be an annoyance when preparing a single layer of atoms, but can also be advantageous for preparing multiple, well spaced trapping planes without requiring additional laser power  \cite{Schlosser2023}. 
Atom array assembly into three dimensional structures is also possible  \cite{Barredo2018,Kumar2018,RLin2025,Kusano2025}. In principle extremely large arrays could be prepared in a single vacuum chamber. Depending on the atom used, the wavelength of the trapping light, and the desired trap depth as little as 1 mW of optical power is required for each site.  Thus 
$10^6$ sites would require 1 kW of optical power. At an atom spacing of $5~\mu\rm m$ the array would cover an area of only 5 mm on a side.  Lasers of this power level are available, although integration with high resolution lens assemblies presents challenges related to thermal management, and mitigation of background light scattering.

\begin{figure}[!t]
\center
\includegraphics[width=.9\columnwidth]{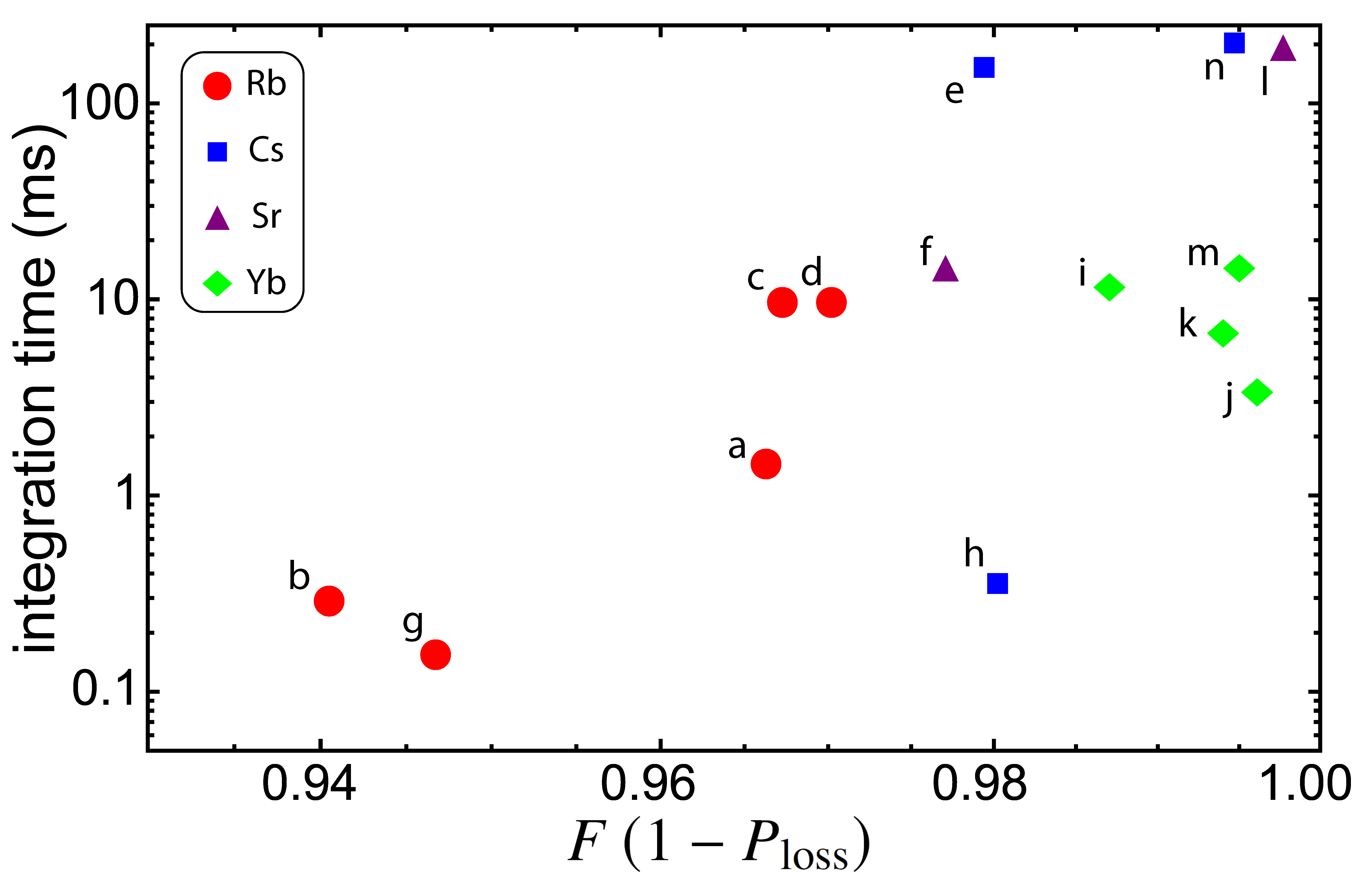}
\caption{\label{fig.measurement} Representative measurements of neutral atom qubit states in free space reported in the literature. The results are quantified in terms of the measurement time and the generalized fidelity given by the state detection fidelity $F$ times the atom retention probability $1-P_{\rm loss}$ . The experiments a,b,g,h  with the shortest integration times used single photon counting detectors and all others used cameras:  
a   \cite{Fuhrmanek2011},
b  \cite{Gibbons2011},
c  \cite{Kwon2017},
d  \cite{Martinez-Dorantes2017},
e  \cite{TYWu2019},
f  \cite{Covey2019a},
g  \cite{Shea2020},
h  \cite{Chow2023},
i  \cite{Huie2023},
j  \cite{Lis2023},
k  \cite{Norcia2024},
l  \cite{RTao2024},
m  \cite{SMa2023},
n  \cite{Scott2025}. }
\end{figure}

An important consideration when preparing very large arrays without defects using atom rearrangement  \cite{WLee2016,  Barredo2016, Endres2016} is the requirement of measuring whether or not an atom is present while retaining the atom for subsequent use after the measurement. In order to prepare an array of $N$ atoms with no defects occupancy  measurements with an accuracy and loss less than $1/N$ are required. Atom occupancy measurements with a fidelity of $F=0.99991(1)$  and a loss probability of $P_{\rm loss}=1-0.99932(8)$ were demonstrated in   \cite{Covey2019a}. These results support preparation of fully occupied arrays with thousands of atoms provided the array assembly time is short compared to the vacuum limited lifetime.

Within quantum circuits it is necessary to measure not just the presence of an atom but the quantum state.   Figure \ref{fig.measurement} presents qubit measurements performed by imaging state dependent fluorescence light. It is notable that the results with the best combination of  fidelity and atom retention   also required the longest integration times. This is because imaging faster requires a higher scattering rate which leads to more heating and atom loss.  High fidelity and very fast measurements are possible using optical cavities  \cite{Bochmann2010,Volz2011, Deist2022, Grinkemeyer2025}, but integration of cavities with large atom arrays remains a challenge.

\section{Rydberg gate fidelity}
\label{sec.fidelity}

\begin{figure}[!t]
\vspace{-.0cm}
\centering
 \includegraphics[width=11.cm]{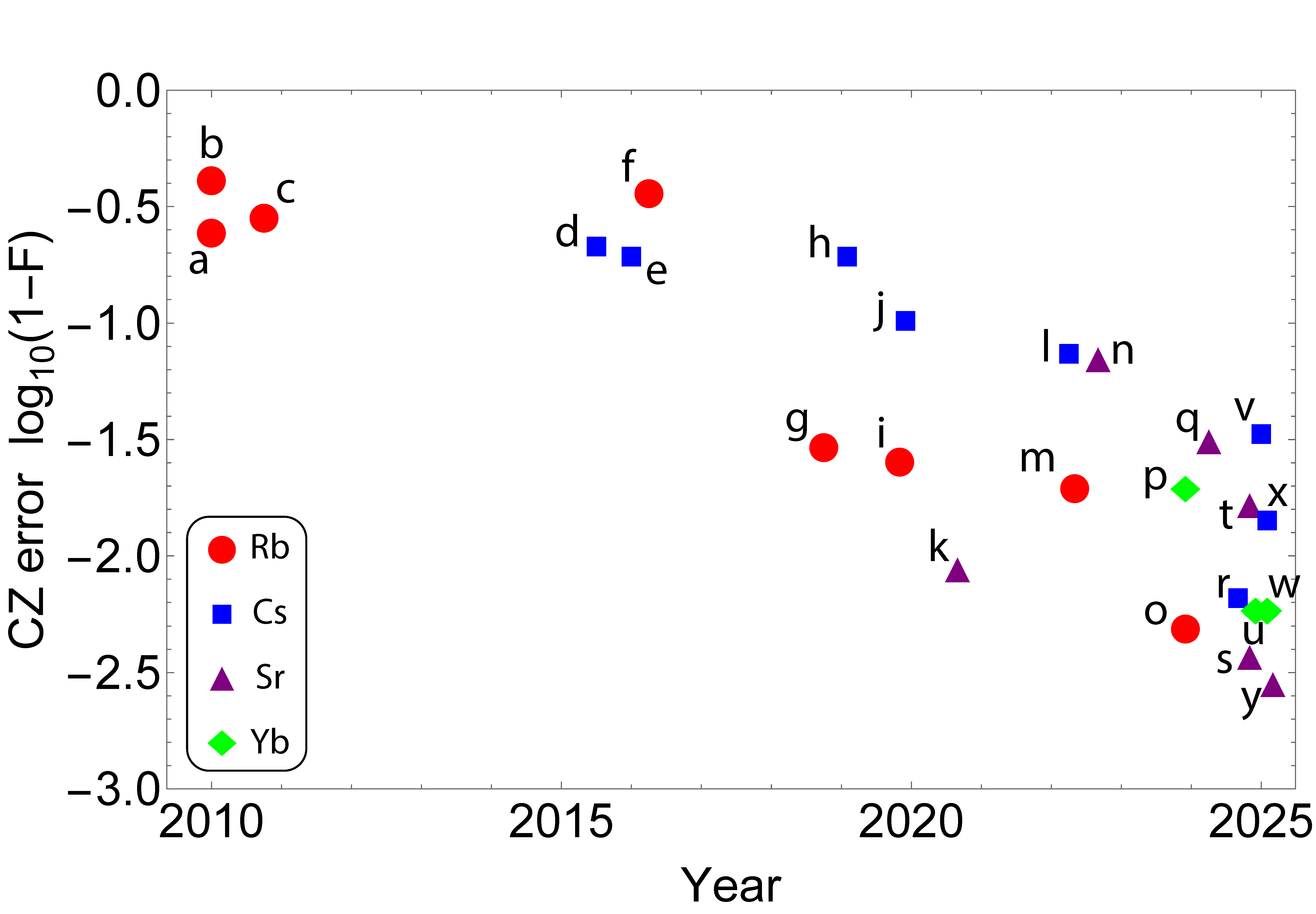}
\vspace{-.0cm}
  \caption{Experimental results for Rydberg $\sf CZ$ gate fidelity. The atomic species are Rb (red circles), Cs (blue squares), Sr (purple triangles), Yb (green diamonds). References: 
    a  \cite{Wilk2010},
  b  \cite{Isenhower2010},
  c  \cite{Zhang2010},
  d  \cite{Maller2015},
  e  \cite{Jau2016},
  f  \cite{YZeng2017},
  g  \cite{Levine2018},
  h  \cite{Picken2019},
  i  \cite{Levine2019},
  j  \cite{Graham2019},
  k  \cite{Madjarov2020},
  l  \cite{Graham2022},
  m  \cite{ZFu2022},
  n  \cite{Schine2022},
  o  \cite{Evered2023},
  p  \cite{SMa2023},
   q  \cite{King2024},
 r  \cite{Radnaev2025},
  s  \cite{Finkelstein2024},
 t  \cite{ACao2024},
 u  \cite{Muniz2025},
v  \cite{MChow2024},
w  \cite{Peper2025},
x  \cite{Chinnarasu2025},
y  \cite{Tsai2025}.
  }
\label{fig.gate_fidelity}
\end{figure}

The original proposals for entangling pairs of atoms and preparing many-body entanglement using the Rydberg blockade mechanism  \cite{Jaksch2000,Lukin2001} were introduced 25 years ago. 
The basic ideas and approaches that have been developed for encoding qubits in neutral atoms, their interactions, and mechanisms for performing and characterizing Rydberg quantum gates have been reviewed in many articles
  \cite{
Saffman2005a, 
Reinhard2007,
Comparat2010, 
Saffman2010, 
Walker2012,
JLim2013,
Marcassa2014,
Saffman2016, 
Browaeys2016, 
Ryabtsev2017,
Saffman2019,
Browaeys2020,
Henriet2020,
Adams2020,
Beterov2020,
XWu2021,
Morgado2021, 
Kaufman2021}.  
First observations of the blockade phenomenon in a many atom ensemble  \cite{Tong2004} were followed by demonstration of blockade between pairs of individual atoms  \cite{Gaetan2009,Urban2009}. Shortly thereafter the first entanglement and two-qubit quantum gates were achieved using Rydberg interactions  \cite{Wilk2010,Isenhower2010,Zhang2010}. The first gate demonstrations had relatively low fidelity and it took some years for the performance to be improved as can be seen in Fig. \ref{fig.gate_fidelity}. The improvement can be ascribed to new developments in several directions which together have resulted in recent rapid progress reaching fidelity well over $99\%$ in several  experiments. 
Notably these high fidelity results were achieved with four different atomic elements $^{87}$Rb  \cite{Evered2023}, $^{133}$Cs  \cite{Radnaev2025}, $^{88}$Sr  \cite{Tsai2025}, and $^{171}$Yb  \cite{Peper2025,Muniz2025}. While a high fidelity two-qubit gate forms the basis for scaling to larger circuits, in the context of optimizing performance for quantum error correction the two-qubit gate fidelity may not directly translate to the fidelity that is achieved for a syndrome extraction protocol or logical operation. In other words optimizing the gate protocol using the logical performance as a cost function may lead to better performance than optimizing the two-qubit gate in isolation.  Several recent papers have addressed this question and predicted improved performance for surface code syndrome extraction  \cite{Jandura2023,Jandura2024b,FQGuo2025}. 

One area of progress has been development of improved gate protocols that reduce the amount of time spent in the Rydberg state, thereby decreasing scattering errors, and which are more robust against experimental imperfections  \cite{Jandura2022, Jandura2023, Pagano2022, Mohan2023}. These time-optimal gates can be designed to work well also without strong blockade  \cite{Poole2025a}, in which case there is some population of doubly-excited Rydberg states during the gate. In this case additional errors arise from inter-atomic forces which must be accounted for  \cite{Robicheaux2021}.

Another factor contributing to improved performance has been the preparation of colder atoms. Early experiments  \cite{Wilk2010,Isenhower2010,Zhang2010} had atom temperatures greater than $50~\mu\rm K$ whereas the recent, high fidelity results were all below $10~\mu\rm K$. Although the Rydberg blockade interaction is intrinsically insensitive to atom temperature there are higher order effects which lead to errors that grow with temperature. In order to prepare entanglement the atoms must be in a superposition of ground and Rydberg states for some amount of time during the gate sequence. Atomic motion, due either to finite temperature, or photon recoil from the laser excitation  \cite{Robicheaux2021}, leads to a stochastic gate phase  as was first pointed out in   \cite{Wilk2010}, which degrades the entanglement fidelity. Furthermore as the temperature increases the trapped atoms become progressively more delocalized which leads to fluctuations in the Rydberg interaction strength, as well as excitation errors due to uncertain overlap with focused Rydberg beams. This latter effect can be reduced, but not eliminated, using shaped beams with a flattened intensity profile  \cite{Reetz-Lamour2008b, Gillen-Christandl2016,Evered2023,Chinnarasu2025}.

A technical aspect of the improvement in gate fidelity that deserves highlighting is the role played by laser noise. Even when the atoms are in their motional ground state phase and intensity noise on the excitation lasers  contribute to imperfect Rabi oscillations between ground and Rydberg states. Although the issue of laser noise was known in the trapped ion and clock communities  \cite{Green2013, Bishof2013, Akerman2015} the specific impact on Rydberg gates with Rabi frequencies in the MHz range had not been studied in detail until the work of Ref.   \cite{deLeseleuc2018}. There it  was shown by numerical simulations that phase noise at the frequency of the ground-Rydberg Rabi frequency was particularly detrimental. Additional analysis established limits on the spectral density of the noise for reaching a desired gate fidelity  \cite{XJiang2023,Day2022}. These limits are not met by the external cavity diode lasers that were typically used for early Rydberg excitation experiments. It was shown convincingly in  \cite{Levine2018} that cavity filtering of the laser phase noise spectrum led to improved control fidelity and recent demonstration of ``spin-locking" diagnostics has enabled direct characaterization of the errors induced by laser noise \cite{Tsai2025}.

With all of these advances the state of the art in Rydberg mediated $\sf CZ$ gate fidelity has reached $\mcF=0.9971(5)$ \cite{Tsai2025}. It is likely that with further experimental improvements $\mcF>0.999$ will soon be achieved. This may occur first in the alkaline earth atoms Sr and Yb which use a one-photon Rydberg excitation pulse starting from the metastable $^3$P$_0$ clock state. 
For the heavy alkali atoms Rb and Cs two-photon excitation from the ground state has been the most widely used approach, although there is also progress on one-photon Rydberg excitation of Cs for implementing an adiabatic dressing   gate \cite{Keating2015,MChow2024}.  With two-photon excitation, high fidelity requires extremely low intensity noise in order to stabilize the differential AC Stark shift between ground and Rydberg states \cite{Maller2015} that is significant when operating at high intensity and large detuning from the intermediate state used for ladder excitation. High intensity and large detuning are requirements in order to have a fast Rabi rate to minimize Rydberg scattering errors, while also minimizing scattering from the intermediate state. 

\begin{figure}[!t]
\center
\includegraphics[width=.7\columnwidth]{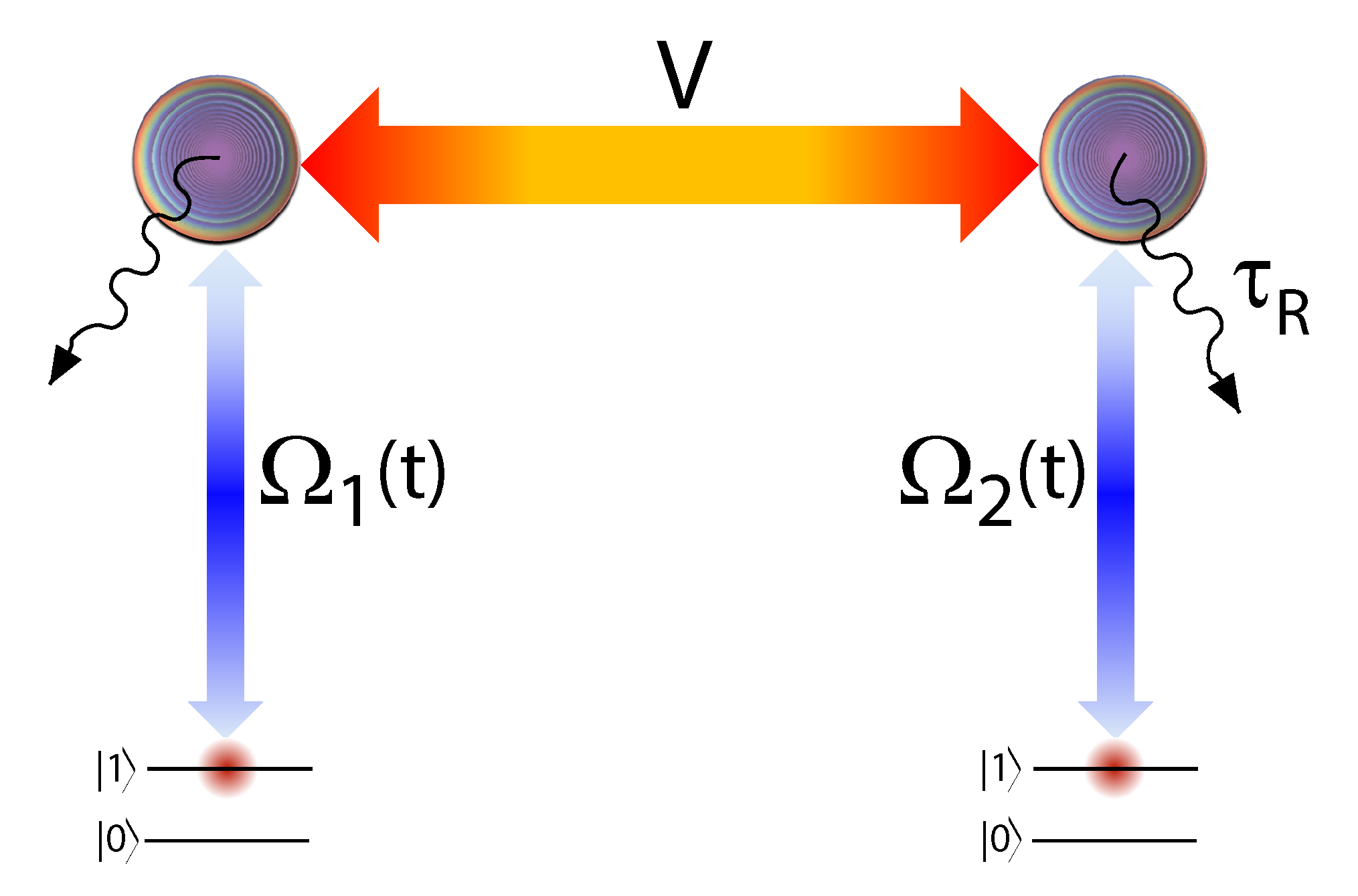}
\caption{\label{fig.gate_limit} Generic Rydberg gate based on excitation of two atoms in a ground state $|1\rangle$  to Rydberg states that interact with energy $V$ and decay to other states with lifetime $\tau_{\rm R}$. The time dependent Rydberg excitation pulses $\Omega_1(t), \Omega_2(t)$ need not be the same for the two atoms.   }
\end{figure}

Irrespective of which atom and laser excitation technique leads to the highest fidelity it is useful to understand the fundamental limits on Rydberg gate fidelity. Consider the setting of Fig. \ref{fig.gate_limit}. Atoms in state $|1\rangle$ are excited to Rydberg states which interact with energy $V$. We assume that due to selection rules or spectral detuning atoms in state $|0\rangle$ are not Rydberg excited.  The pulse profiles $\Omega_1(t), \Omega_2(t)$ are designed such that after a total time $T$ all population returns to the ground state, neglecting any spontaneous decay to other states. In the interval $0\le t\le T$ there is some Rydberg population $P_1$ of atom 1 alone, $P_2$ of atom 2 alone, and $P_{12}$ of both atoms simultaneously excited. This gives an integrated Rydberg population of 
\begin{equation}
    P_{\rm R}=\int_0^Tdt\, P_1(t)+P_2(t)+2P_{12}(t).
\end{equation}
It was shown in  \cite{Wesenberg2007} assuming the atomic qubits start in a separable pure state, and considering the growth of the von Neumann entropy during the gate, that in order to create one unit of entanglement we require 
\begin{equation}
P_{\rm R}\ge \frac{2}{V}.
\end{equation}
This argument is not specific to Rydberg interactions and holds for any qubits that are tranferred to interacting states in order to prepare entanglement, 
Assuming further that spontaneous emission leads to an error $\epsilon$ we may write 
$\epsilon=P_{\rm R}/\tau_{\rm R}$ with $\tau_{\rm R}$ the Rydberg lifetime so 
\begin{equation}
\epsilon_{\rm min}\ge \frac{2}{V\tau_{\rm R}}.
\label{eq.Vtau}
\end{equation}
Equation (\ref{eq.Vtau}) sets a lower bound on the achievable infidelity of a Rydberg $\sf CZ$ gate, but it does not tell us how to design $\Omega(t)$ to saturate the bound. 

It is instructive to estimate this bound using atomic parameters and compare with known gate protocols. As an example consider two Cs atoms excited to the $|68p_{3/2}, m_j=3/2\rangle$ state that are separated by $R=2.5~\mu\rm m$ with the interatomic axis aligned along the quantization axis. The Rydberg intgeraction energy computed with the Pair interaction code \cite{Weber2017} is $V=2\pi\times 215 ~\rm MHz$ and the Rydberg lifetime in a room temperature environment is \cite{Beterov2009,Beterov2009b} $\tau_{\rm R}=198~\mu\rm s$. These values give 
\begin{equation}
\epsilon_{\rm min}\ge 7.5\times 10^{-6}.
\end{equation}
While impressive, this bound should not be considered a lower limit as higher Rydberg states with longer lifetimes would further reduce the error floor. 

There is an important open question of how to design $\Omega(t)$ to reach this bound.  The original Rydberg $\pi-2\pi-\pi$ gate proposal \cite{Jaksch2000} has an error scaling \cite{XZhang2012} $\epsilon\sim1/(V\tau_{\rm R})^{2/3}$ which is found from optimizing the Rabi frequency to balance errors between imperfect rotations at finite interaction strength ( $\sim|\Omega/V|^2$)  and the Rydberg scattering error $(\sim 1/\Omega\tau_{\rm R}) .$ It turns out that the $\pi-2\pi-\pi$ gate can be modified to have an error scaling as $1/(V\tau_{\rm R})$ \cite{Cole2023}, and also other protocols that rely on individual addressing of control and target qubits reach the optimum $1/(V\tau_{\rm R})$ scaling \cite{Petrosyan2017}.  

Nevertheless, protocols that have $\Omega_1(t)\ne\Omega_2(t)$ so the temporal waveforms are different on the two qubits require more complex control systems. Adiabatic proposals for gates with symmetric qubit control $(\Omega_1(t)=\Omega_2(t))$ have been proposed \cite{YSun2020,Saffman2020}, and the qubit symmetric, two-pulse protocol from \cite{Levine2019} was the first to demonstrate fidelity above  97\%. The time-optimal protocols \cite{Jandura2022,  Pagano2022, Mohan2023} which have delivered the highest gate fidelities are qubit symmetric but do not rely on adiabaticity to return all initial states to the ground state at the end of the gate. Instead they rely on fast driving to minimize time spent in the Rydberg state with a time dependent phase profile that ensures that initial states with both one- and two-atoms Rydberg coupled completely return to the ground state. 

In the limit of strong blockade the protocols that come closest to the limit set by Eq. (\ref{eq.Vtau}) are the two-atom dark state gate \cite{Petrosyan2017} which reaches $\epsilon/\epsilon_{\rm min}\simeq 19$, the time-optimal gate \cite{Jandura2022} which reaches $\epsilon/\epsilon_{\rm min}\simeq 15$ and the modified $\pi-2\pi-\pi$ gate \cite{Cole2023}. 
With strong blockade additional errors due to simultaneous Rydberg excitation of both atoms, which leads to inter-atomic forces, are negligible. In the limit of weak blockade modified time-optimal profiles were found to reach  $\epsilon/\epsilon_{\rm min}= 2.1$ \cite{Poole2025a}. Including two-atom interaction dephasing and the effect of photon recoils the predicted error increases to $\epsilon/\epsilon_{\rm min}= 3.0$. Photon recoil heating can be minimized using more complex time-dependent pulses \cite{Lis2023,ZZhang2024,vanDamme2025}, but this has not yet been demonstrated in practice with Rydberg gates.

\section{The connectivity challenge}
\label{sec.connectivity}

Running complex quantum algorithms on a digital quantum computer requires gate operations involving qubits that may not be next to each other. Any physical interaction between qubits has a finite range. While there are settings for which all qubits are interconnected, such cases do not support scaling to arbitrarily large qubit numbers.   As an example  consider coupling many atomic qubits to the mode of an optical resonator such that photons mediate all to all connectivity \cite{Pellizzari1995,Vaidya2018,Ramette2022}.  Even though  photon mediated coupling does not intrinsically have a limited range there is a practical limit to how many atoms can be strongly coupled to a single cavity.

For quantum computation without error correction the additional gate operations needed to
connect non-proximal qubits add operational overhead and reduce the useful computational depth. The severity of establishing connectivity depends on the algorithm being run, and can be reduced with optimized compilation that is aware of the physical level hardware capabilities \cite{DBTan2024,Schmid2024}. There are also encoding techniques for specific problems that effectively provide long-range interactions at the cost of consuming more physical qubits \cite{Lechner2015, Nguyen2023}.
A well designed  error corrected processor that operates with logical qubits should have a computational depth that greatly exceeds that of a NISQ machine.  Nevertheless, useful error correction does not imply an arbitrarily small logical error rate, just an error rate that is smaller than that of the underlying physical operations. Thus also logical processors will suffer reduced useful circuit depth due to resources required for connectivity, as well as longer operation times. 

In this section we compare several approaches to establishing non-local connectivity for logical qubits in atom arrays with an emphasis on the execution time requirements. Performing a gate between 
neighboring logical qubits
may require $\it nonlocal$ connectivity at the physical level when the gate is based on coherent interactions. An important alternative is lattice surgery which enables gates between logical qubits, requiring only local physical gates plus measurements and auxiliary qubit resources \cite{Horsman2012}.  
To focus the discussion we will assume encoding in the  $[[d^2,1,d]]$  rotated surface code. This encoding requires $2d^2-1$ physical qubits for each logical qubit assuming no reuse of ancillas between $\sf X$ and $\sf Z$ stabilizer extractions. Based on  a standard depolarizing noise model the logical error rate of the rotated code with a minimum-weight perfect-matching decoder is \cite{ORourke2025}
\begin{equation}
    p_{\rm L}=0.08(p/p_{\rm th})^{0.58 d-0.27}.
\end{equation}
where the  threshold physical error rate is $p_{\rm th}=0.0053.$ 
Taking $\mcF=0.9992$ or $p=0.0008$, which appears within reach of Rydberg gates, a code distance of $d=10$ is sufficient to reach within about a factor of two  of $1/p_{\rm L}=10^6$. 
At this code distance 100 logical qubits would require 19900 physical qubits, plus the overhead of magic state distillation for non Clifford operations. 
Such a ``Megaquop" machine with 100 error corrected logical qubits is likely to be of great  interest from a scientific perspective \cite{Preskill2025}. 
While atom arrays are currently one of the most promising approaches for reaching this combination of low error rates and high qubit counts the clock rate of neutral atoms is slow compared to other platforms. With that challenge in mind the following sections provide a comparison of logical gate times based on several different approaches.

\subsection{long-range Rydberg interactions}
\label{sec.long_range}

The rotated surface code has $[[n,k,d]]=[[d^2,1,d]]$ and the gate time for a transversal logical $\sf CZ$ using sequential long-range Rydberg interactions is $t_{\rm gate}\le d^2(t_{\sf CZ}+t_{\rm beam})$ where $t_{\sf  CZ}$ is the time for the Rydberg gate and $t_{\rm beam}$ is the time to point the optical control beams at a new atom pair. This is a strict equality for sequential gate operations and can be quicker if some of the atom pairs are far enough separated to allow for parallel gate operations (see  \cite{Poole2025a} for an analysis of the separation required to minimize interference effects between gate pairs). Significantly, the ability to implement $\sf CZ$ gates transversally, in contrast  to the lattice surgery methods discussed in Sec. \ref{sec.surgery}, allows for efficient decoding with a reduced number of measurement rounds \cite{HZhou2025}. 

\begin{figure}[!t]
\center
\includegraphics[width=1.\columnwidth]{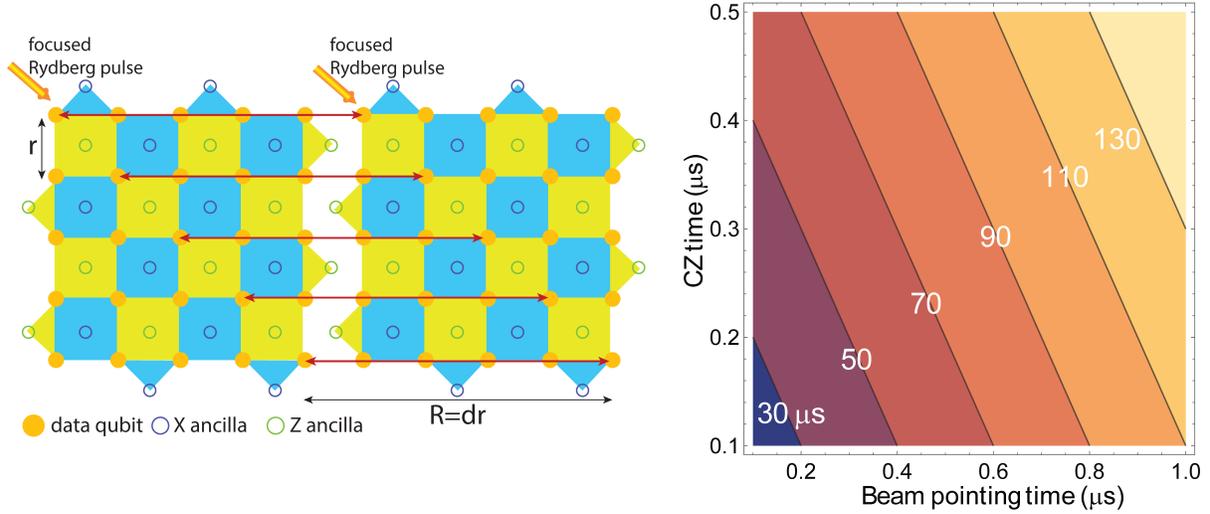}
\caption{\label{fig.parallel_gate} Time for a transversal logical $\sf CZ$ gate that is implemented sequentially  for a $d=10$ rotated surface code as a function of the physical $\sf CZ$ time and the optical beam pointing time.  The graphic on the left indicates a subset of the $d^2$ long-range gates needed to implement the transversal $\sf CZ$.  }
\end{figure}

Figure \ref{fig.parallel_gate} shows the logical gate time assuming $d=10$. The calculated times which can be well under $100~\mu\rm s$ assume two-qubit Rydberg gates with an optical control system that excites pairs of atoms using dynamically scanned, tightly focused beams \cite{Graham2022,Radnaev2025}. The achievable gate speed is strongly dependent on the performance of the optical beam scanning device. Acousto-optic deflectors are compatible with the sub-microsecond times assumed in the figure, whereas other technologies such as spatial light modulators and digital mirror arrays operate at much slower speeds. The combination of these approaches may allow fast reconfiguration together with the ability to address large qubit arrays \cite{Graham2023}.  

\begin{figure}[!t]
\center
\includegraphics[width=.9\columnwidth]{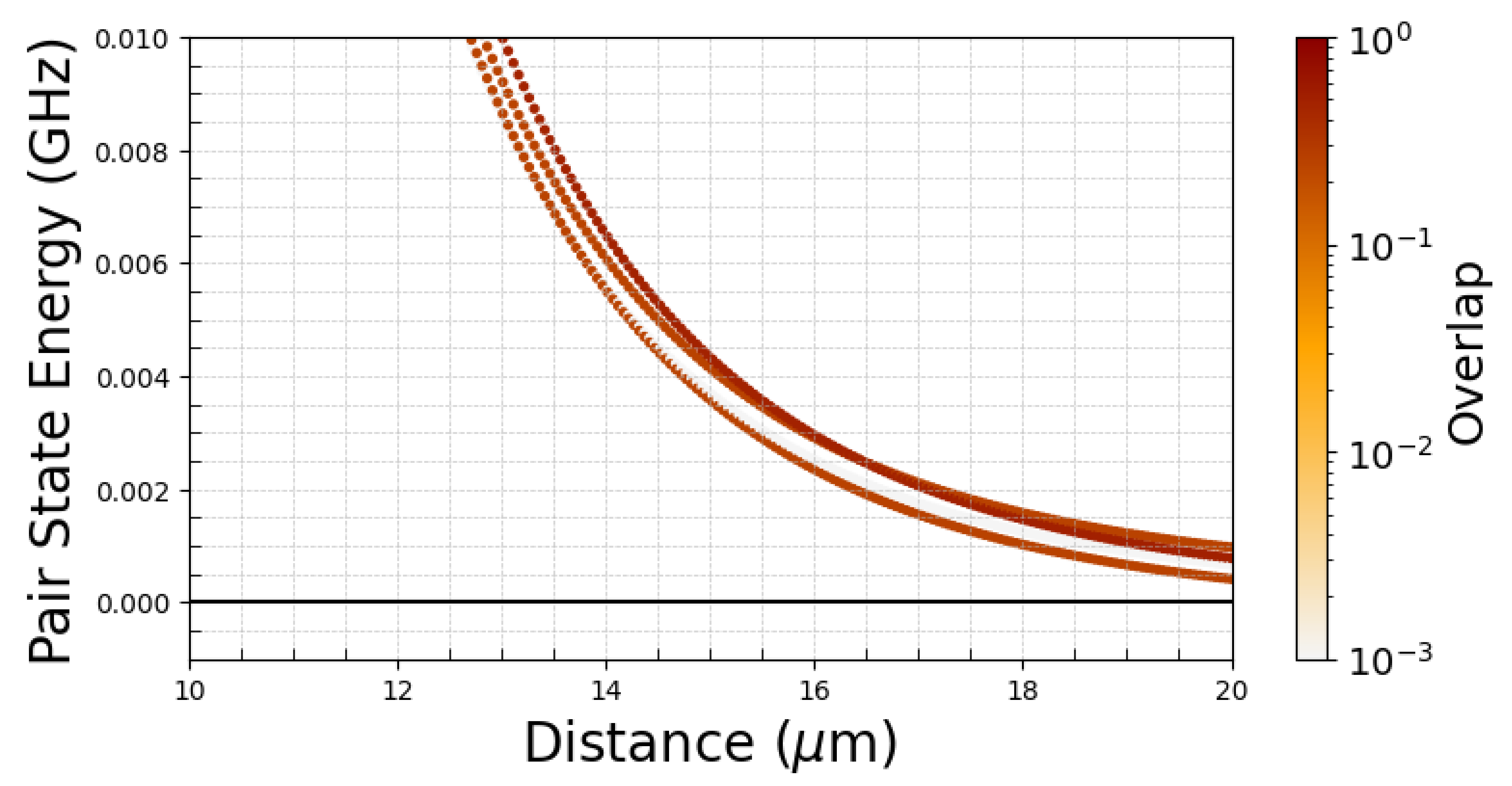}
\caption{\label{fig.interaction} Rydberg interaction strength for Cs atoms excited to $100s_{1/2},m_j=1/2$ states quantized along the atomic separation axis ($\hat z$)  with a magnetic field of $B_z=10^{-5}~\rm T$ .  Calculation performed with pair interaction code \cite{Weber2017}.   }
\end{figure}

A limitation of this approach is the finite range of the Rydberg interaction. At $d=10$ and a tight spacing between physical qubits of $r=1.4~\mu\rm m$ the required interaction distance is $R=dr=14~\mu\rm m$ which is beyond the range of demonstrated Rydberg gates. A high fidelity gate at this distance appears feasible by combining progress in several directions. The use of highly excited Rydberg states can provide close to $V=2\pi\times 6~\rm MHz$ of interaction strength at  this distance as shown in Fig. \ref{fig.interaction}. With this value for $V$ a modified time-optimal gate profile that provides high fidelity with modest interaction strength can reach $\mcF=0.9992$ with a gate time of $0.46~\mu\rm s$ \cite{Poole2025a}. With an optical beam scanning time of $0.5~\mu\rm s$ this leads to a logical gate time of about $90~\mu\rm s$.

Extending this approach to larger code distance which implies larger $R$ appears challenging. Our assumption of  $r=1.4~\mu\rm m$ atomic spacing cannot be significantly reduced without running in to the limit set by the wavefunction of a Rydberg excited electron  overlapping with a neighboring ground state atom.  The Rydberg interaction strength scales as $1/R^3$ at short distances (limit of resonant dipolar interactions) and $1/R^6$ at long-range(van der Waals limit)  \cite{Walker2008}. Judicious application of dc or microwave fields can be used to reduce the F\"orster defects and extend the $1/R^3$ scaling to longer distances \cite{Bohlouli-Zanjani2007, Ravets2014, Beterov2016b, Kurdak2025}.

It is perhaps interesting to check how large the interaction could ideally be at say $R=20~\mu\rm m$ which could enable $d=15$ and 100 Megaquop performance. 
At F\"orster resonance the interaction of two atoms excited to Rydberg states that interact through a single channel can be expressed as $V(R)=-C_3 D^{1/2}/R^3$  where $D$ is an angular factor depending on the interaction channel and 
\begin{equation}
    C_3=\frac{e^2}{4\pi\epsilon_0}\frac{\langle \gamma_\alpha||r_a||\gamma_a\rangle\langle \gamma_\beta||r_b||\gamma_b\rangle}{\sqrt{2j_\alpha+1}\sqrt{2j_\beta+1}},
\end{equation}
where $e$ is the electronic charge, $\epsilon_0$ is the permittivity of free space, the reduced matrix elements are calculated in the fine structure basis, and $\gamma$ denotes the quantum numbers of the initial Rydberg levels $a,b$ that are coupled to levels $\alpha,\beta$ through a  F\"orster interaction \cite{Walker2008,Beterov2015}. Consider as an example Cs $a,b$ states $n_as_{1/2}, n_b s_{1/2}$  interacting with $n_\alpha p_{1/2}, n_\beta p_{1/2}$ states through the triplet channel with $m=m_a+m_b=0$. In this case $D=16/9$ and with $n_a=99, n_b=100, n_\alpha=98, n_\beta=100$ we find 
\begin{equation}
    V(R)/h= -\frac{44,000}{R^3}~ (\rm MHz)
\end{equation}
with $R$ in units of $\mu\rm m$. Thus, using $n=100$ Rydberg states at $R=20~\mu\rm m$ will give at most about $5~\rm MHz$ of interaction strength. This is roughly a factor of five larger than the undressed interaction shown in Fig. \ref{fig.interaction}.

There is another potential path to achieving longer range interactions by introducing additional atoms to provide a quantum interconnect bus. Suppose we wish to achieve a strong interaction between two atoms separated by a large distance $R$. In the van der Waals limit the interaction scales as $V=C_6/R^6$, with $C_6\sim C_3^2/\delta$ where $\delta$ is the F\"orster energy defect. If an additional atom is inserted midway in between the Rydberg interaction with the original atoms wil be
\begin{equation}
    V'=\frac{C_6}{(R/2)^6}=64 V.
\end{equation}
By judicious design of Rydberg pulse profiles it is possible to achieve an effective interaction between the original atoms that is much greater than $V$, although not as large as $V'$ \cite{Cesa2017,YSun2024,Doultsinos2025}.

We  conclude that transversal logical gates with the surface code based on native long-range Rydberg interactions will not  scale beyond a code distance of $d=10$ using the surface code, and even that will be extremely challenging to reach. There are methods to increase the Rydberg interaction range based on reduction of the F\"orster defect, or on adding additional atoms to mediate the interaction. Such methods may  extend the interaction range beyond $20 ~\mu\rm m$ which would make a code distance of $d=15$ possible. An  alternative path to higher code distance is to use quantum low density parity check (qLDPC) codes that have non-local check operations and a higher code rate $k/n$ than the surface code. Implementation of high rate codes for quantum memory with atom arrays has been analyzed in several recent papers \cite{Viszlai2023, QXu2024, Poole2025a, Pecorari2025}. An optimized neutral atom implementation \cite{Poole2025a} of the  bicycle code introduced in  \cite{Bravyi2024}  identified a qubit layout for a $d=18$ code that is realizable with long-range Rydberg gates, without invoking atom motion or any improvement from Rydberg dressing.

\subsection{Transport based gates}
\label{sec.transport}

An alternative approach which enables arbitrary longer range connectivity is qubit motion during circuit operation. The concept of relying on physical motion of qubits dates back to early ideas for scaling trapped ion quantum processors \cite{Cirac2000,Kielpinski02} and has been implemented with great success in recent trapped ion demonstrations \cite{Pino2021,Moses2023}. In trapped ion platforms, and also  extensions to shuttling of quantum dots in semiconductor structures \cite{Flentje2017,Zwerver2023}, motion relies on applying a sequence of voltages to closely spaced electrodes to provide a time dependent force on the charged particle. A drawback of the trapped ion approach, in addition to the complexity of controlling many electrodes, is that motion induces substantial heating of the center of mass motional state so recooling is necessary as part of the control sequence.

Neutral atoms trapped in optical potentials defined by light can be transported much more efficiently in terms of the number of required control signals: either by varying the phase of a single optical beam to continuously move the nodes of an interference pattern \cite{Mandel2003b}, or by scanning an optical tweezer using a varying frequency applied to an acousto-optical deflector \cite{Beugnon2007}. When applied to moving arrays of atoms in parallel the optical tweezer approach becomes a powerful technique for connecting logical qubits in a neutral atom processor \cite{Bluvstein2022,Bluvstein2024,Reichardt2024}.

We may compare the time to implement a logical {\sf CZ} gate using atom motion with the case of long-range Rydberg interactions discussed in Sec. \ref{sec.long_range}. The attainable transport speed is limited by the need to minimize motional heating \cite{Hickman2020, Hwang2025}. Several different transport profiles have been demonstrated including constant jerk \cite{Bluvstein2022}, and minimal jerk \cite{LLiu2019,Matthies2024,Finkelstein2024}, where jerk is the time derivative of acceleration \cite{Hogan1984}.  Remarkably it is also possible to throw and catch atoms over distances of several $\mu\rm m$ \cite{Hwang2023}.

For the minimal jerk profile we find using the analysis in  \cite{Carruthers1965} that the time to move a distance $R$ while incurring $\delta n$ quanta of motional excitation is \cite{Chinnarasu2025} 
\begin{equation} 
t_{\rm mj}=\frac{2^{1/2}15^{1/3}R^{1/3}}{\delta n^{1/6}x_{\rm ho}^{1/3}\omega_0}.
\label{eq.tmj}
\end{equation}
Here $x_{\rm ho}=(\hbar/(2m\omega_0))^{1/2}$ is the harmonic oscillator length with $\hbar$ the reduced Planck constant, $m$ the atomic mass, and $\omega_0$ the trap vibrational frequency. 
For a wide range of parameters the minimal jerk profile provides less motional heating than constant jerk for the same transport time.

\begin{figure}[!t]
\center
\includegraphics[width=1.\columnwidth]{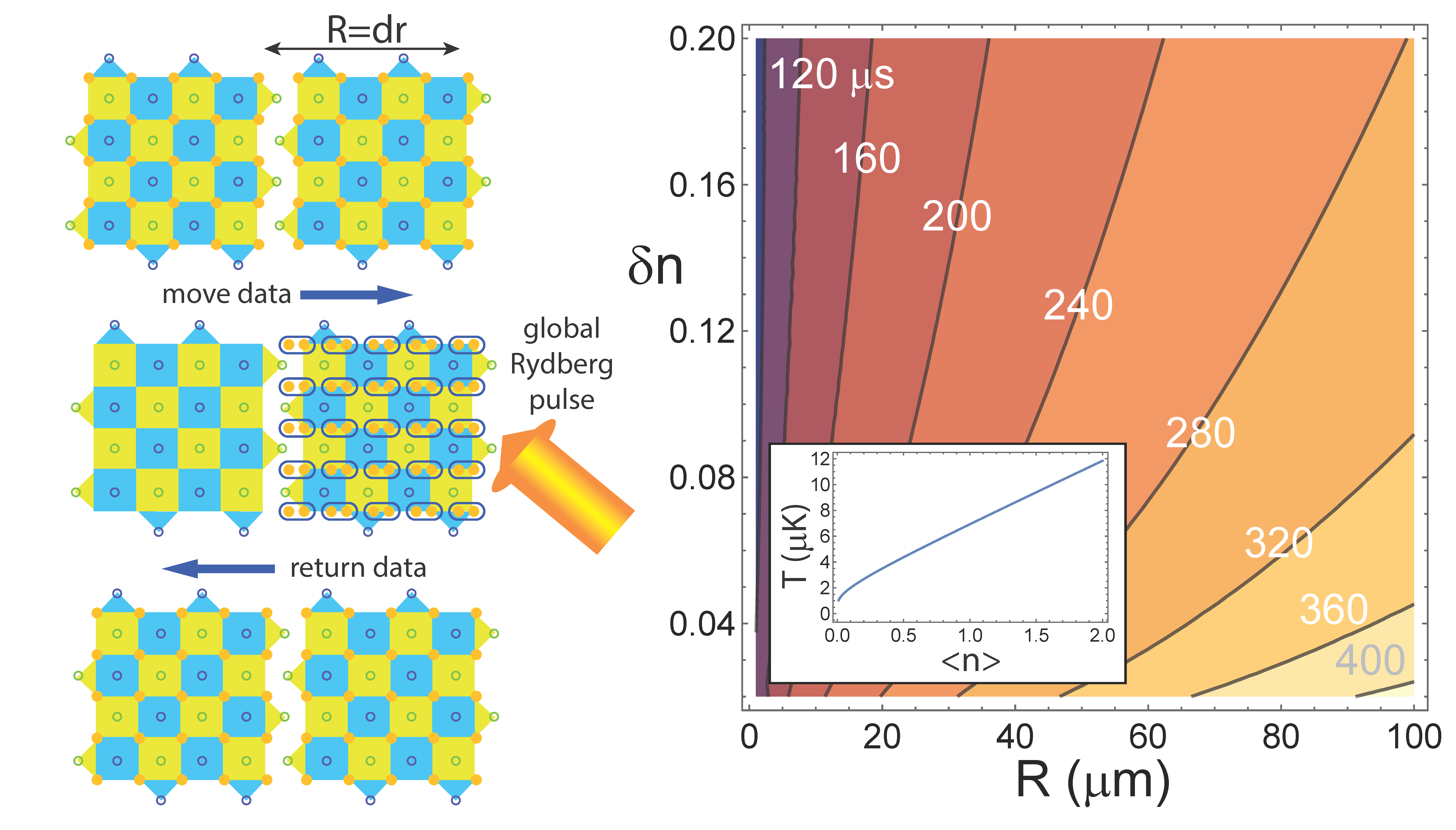}
\caption{\label{fig.transport_gate}  Sequence for a transport based logical $\sf CZ$ involves (top to bottom) moving the data qubits a distance $R$, applying a Rydberg pulse, and then moving the data qubits back. The contour plot shows the minimal jerk time to implement the move back and forth, neglecting any additional time for the Rydberg pulse, as a  function of the distance $R=dr$, and the round trip motional heating incurred. The inset shows the one-dimensional atomic temperature as a function of the mean number of motional quanta assuming a trap frequency of $\omega_0=2\pi\times 100 ~\rm kHz$ and the expression $k_{\rm B}T=\hbar\omega_0/\ln(1+1/\langle n \rangle)$.   }
\end{figure}

The sequence for a transport enabled logical gate is illustrated in Fig. \ref{fig.transport_gate} with the indicated time based on the minimal jerk profile and not accounting for any additional time for the Rydberg pulses. With sufficient laser power all $d^2$ gates for a distance $d$ surface code can be implemented in one time step. The laser power requirement could be reduced by addressing $d$ pairs in each row in each time step, with $d$ time steps in total. Even at $d=20$ using $d$ time steps would add less than $10\%$ to the values shown, assuming gate and beam pointing timings corresponding to the middle of the range shown in Fig. \ref{fig.parallel_gate}. 

Parallel implementation of a transport based transversal gate relies on positioning the atom pairs at a small spacing, relative to the distance to neighboring pairs. Doing so allows multiple gate pairs to be performed in the same time step. As long as the array pitch $r$, which is the distance between pairs, satisfies $r\gg r_g$, with $r_g$ the distance between atoms in one gate pair there will be minimal crosstalk.  The sensitivity of the time-optimal $\sf CZ$ gate to interaction crosstalk was studied in  \cite{Poole2025a} where it was found that provided $\eta_i=\sum_j\Delta_{ij} t_{\rm gate}/(2\pi)<0.01$ the gate infidelity due to crosstalk is upper bounded by $10^{-4}$. Here $\Delta_{ij}$ is the energy shift of Rydberg atom $i$ due to all other atoms $j$, and $t_{\sf CZ} $ is the gate time.  In the van der Waals interaction limit the shifts scale as $\Delta_{ij}\simeq \frac{V(r_g)}{\hbar} \left(\frac{r_g}{r_{ij}}\right)^6.$ For a high fidelity time-optimal gate we require \cite{Jandura2022} $\Omega t_{\sf CZ}/\hbar\ge 7.6$ with $\Omega$ the Rydberg excitation Rabi frequency. Using $(V/\hbar)/\Omega=20$ which is adequate for a high fidelity gate and approximating $\eta_i$ by the worst case of $\eta\simeq4 \Delta(r) t_{\sf CZ}/(2\pi)$ we find 
\begin{equation}
    r\ge 4.6 r_g. 
\end{equation}
Using $r_g=1.4~\mu\rm m$ (the same tight array spacing as was assumed in Fig. \ref{fig.parallel_gate}) we require 
$r=6.44~\mu\rm m$ so the transport distance for a $d=10$ code is $R=64~\mu\rm m$. Referring to Fig. 8 the total transport time at a heating increment of $\delta n=0.1$ is $270~\mu\rm s$. Comparison with the time for  a long-range Rydberg gate shown in Fig. \ref{fig.parallel_gate} depends on assumptions about the beam pointing time. With sub $\mu\rm s$  beam pointing the motion enabled gate is slower by a factor of about 2 - 5.

\subsection{Lattice surgery}
\label{sec.surgery}

A third approach to implementing gates on pairs of logical qubits is lattice surgery \cite{Horsman2012}. In order to perform a $\sf CNOT$ gate between two surface code qubits of distance $d$ an auxiliary patch of surface code is utilized. A sequence of $d$ physical gates and $d$ measurements is performed along the boundary of the first logical qubit and the auxiliary region. This merging operation creates a code patch of dimension $d\times 2d$. The merged patch is then split with a sequence of $d$ gates and $d$ measurements between it and the second logical qubit. The net result after $2d$ physical gates and $2d$ measurements is a $\sf CNOT$ gate between the two logical qubits. 

A feature of lattice surgery is that all of the operations occur along the boundaries of the logical patches which eliminates the need for long-range gates or long distance motion. The time for the  operation if it is performed in place using individual addressing beams that are scanned is $ t_{\rm ls}=2d (t_{\sf CZ}+t_{\rm beam}+t_{\rm meas})$, with $t_{\rm meas}$ the measurement time (the time for the additional Hadamards to convert $\sf CZ$ to $\sf CNOT$ is negligible).  If the operation is performed by moving the atoms so that there are $d$ closely spaced pairs suitable for a global $\sf CZ$ gate the time is $t_{\rm ls}=2 t_{\sf CZ}+4 t_{\rm move}+2d t_{\rm meas}.$ The required move distance is a single lattice period $r$. Using the value from the previous section and setting $\delta n=0.1$ for the sequence ($\delta n=0.1/4$ per move) we find $t_{\rm move}=71~\mu\rm s$. For a code distance of $d=10$ or more the lattice surgery time is dominated by the measurement time. Referring to Fig. \ref{fig.measurement} the  high fidelity measurements which are well below the surface code threshold have $t_{\rm meas}$ of at least a few ms. 
We see that lattice surgery is not competitive with either long-range Rydberg or transport based gates. Substantial improvements in measurement time while maintaining high fidelity would be needed to change this conclusion.

\subsection{Connectivity comparisons}

The previous three sections compared the time needed for $\sf CZ$ gates between surface code logical qubits with code distance $d=10$ using three different approaches.  
The fastest option is long-range Rydberg interactions, although this does not scale beyond $d=10$ without enhancement of the Rydberg interaction range, which is possible, but challenging. Even at $d=10$ an in-place logical gate is challenging due to the requirement on very small atom spacing within each logical qubit. Achieving high fidelity under these conditions will require high performance optics, capable of producing very small focal spots with low light scattering. For smaller distance codes, say $d=5$, the atom spacing could be increased to a more relaxed $r\sim 2~\mu\rm m$ but with our assumed gate fidelity of $\mcF=0.9992$ the achievable circuit depth would be limited to  
$1/p_{\rm L}=1800$, far short of the Megaquop goal. 

The lattice surgery approach is relatively straightforward with current technology, either based on 
in place Rydberg gates or atom transport. The drawback is the slow speed due to the measurement time bottleneck. This leaves the transport based gates described in Sec. \ref{sec.transport} as the fastest approach based on current experimental capabilities. The timings indicated in Fig. \ref{fig.transport_gate} are for a gate between neighboring surface code patches. Assuming a move distance of $64~\mu\rm m$ and a heating increment of $\delta n=0.1$ motional quanta we found a logical gate time of $t_{{\rm L}, \sf CZ}= 270~\mu\rm s$.

Let us proceed by estimating the logical gate time averaged over a $10\times 10$ array of 100 logical qubits. The average Manhattan distance between two sites in a $10\times 10$ array is 6.67, in units of logical qubit size.
Since the minimal jerk time 
in Eq. (\ref{eq.tmj}) scales as $R^{1/3}$ we need to find the average of the Manhattan distance to the $1/3$ power
which is 1.82. An estimate for the array averaged gate time is thus
$$
\left\langle t_{{\rm L}, \sf CZ}\right\rangle= 1.82\times 270 = 490 ~\mu\rm s.
$$
It is interesting to compare this with the geometry of Ref.  \cite{Bluvstein2024} which
moved qubit pairs to be entangled to a common zone located on the edge of the qubit array. This has the benefit that the Rydberg beams 
point to a fixed location and do not have to be scanned. The average Manhattan distance to the  $1/3$ power to move to a location just outside the middle of one edge of the array is 1.96, giving $\left\langle t_{{\rm L}, \sf CZ}\right\rangle= 530 ~\mu\rm s$ which is only slightly larger.  In either case we see that the cycle time for the $d=10$, 100  logical qubit array will be at best about a kHz.

\section{Syndrome extraction and mid-circuit measurements }
\label{sec.midcircuit}

So far we have discussed operational times for implementing gates between logical qubits. In an error corrected logical processor a large fraction of the operational time will be spent checking idling qubits for errors due to decoherence, and checking for errors arising during gate operations. 
The most widely used approach for measuring neutral atom qubits is state selective light scattering. A simple example is provided by hyperfine encoding  of a qubit in an alkali atom. The ground $ns_{1/2}$ electronic state has two hyperfine levels with total angular momentum $f_\pm=I\pm 1/2,$ with $I$ the nuclear spin. The transition $|ns_{1/2},f_+\rangle \rightarrow |np_{3/2},f_++1\rangle$ is cycling and can be used to scatter many photons to detect occupation of the $f_+$ level. The $f_-$ level is detuned by the ground hyperfine splitting which is 6.8 GHz in $^{87}$Rb, 9.2 GHz in $^{133}$Cs, and thereby sufficient to render the $f_-$ level dark to the probing light.  

Two problems arise with this type of measurement. First, although the transition out of the $f_+$ level is cycling, there is a non-zero Raman rate for scattering from $f_+\rightarrow f_-$ via the $|np_{3/2,}f=f_+\rangle$ state (e.g. $|6s_{1/2},f=4\rangle \rightarrow |6p_{3/2},f=4\rangle\rightarrow |6s_{1/2},f=3\rangle$ in Cs). To prevent this happening it is possible to measure the state using two $\sigma_+$  polarized beams that drive the transition $|ns_{1/2},f_+,m=f_+\rangle \rightarrow |np_{3/2},f_+ +1,m=f_++1\rangle$ which is cycling and does not suffer from Raman transitions. This works well but leads to transverse heating of the atom since the readout and cooling beams are only applied along a single axis. To mitigate heating loss it is necessary to use relatively deep trap potentials which requires more trap laser power. Details concerning relevant experimental parameters can be found in several papers \cite{Gibbons2011, Fuhrmanek2011, Jau2016, Kwon2017, Martinez-Dorantes2017, Martinez-Dorantes2018,Chow2023}. 

The second issue is that for quantum error correction based on qubit measurements for syndrome extraction it  is necessary to measure ancilla qubits without disturbing the state of proximal data qubits. The standard measurement approach described above does not fare well because the data qubits act as resonant absorbers with a cross section $\sim \lambda^2$ but are only spaced by a few $\lambda$. In order to avoid measurement crosstalk several approaches have been developed based on either transport to a distant measurement zone \cite{Deist2022, Bluvstein2024}, shelving of data qubits in dark states or hiding of data qubits \cite{Graham2023b,Lis2023, Norcia2023, SMa2023, Bluvstein2025}, hiding combined with cavity assisted fast readout \cite{BHu2025}, or arrays with two atomic isotopes or species \cite{Beterov2015, CSheng2022, Singh2023,CFang2025}. There are also proposals based on electromagnetically induced transparency which have not yet been demonstrated \cite{Giraldo2022, Saglam2023} .

From the perspective of speed and fidelity of operations almost all of these techniques require some extra operations, beyond what is required for a state selective measurement without regard to crosstalk. The extra operations are either transport to a cavity or a measurement zone which adds time, or the need to shelve or hide atoms that are not intended to be measured by applying additional control pulses, which adds errors from imperfect operations. The one exception is the two-species approach. When seen as part of a  syndrome extraction circuit for error correction the operations required for two-species are the same as would be required when using a single species for data and ancilla qubits. The only significant difference is that the entangling gate is applied between atoms of different types. There are F\"orster resonances available which make, for example, the interspecies Rb-Cs interaction very strong \cite{Beterov2015,Ireland2024}. Similar resonances can be found in other atomic combinations such as Rb-K \cite{Samboy2017,Otto2020}. For these reasons a two-species architecture appears very promising although the demonstrated Rydberg entanglement fidelity \cite{YZeng2017,Anand2024} has not yet reached the same level as for single species gates. 

An additional feature of the two-species approach is that the ancillae, which are repeatedly measured, can be recooled without disturbing quantum information stored in the data qubits \cite{Singh2023}. Furthermore, by mapping the data qubit states onto ancilla qubits, the data atoms can also be recooled inside a long calculation, which can then continue with the roles of data and ancilla atoms swapped, or the information can be swapped back. This mid-circuit cooling capability appears to be uniquely straightforward in a two-species architecture.

\begin{figure}[!t]
\center
\includegraphics[width=1.\columnwidth]{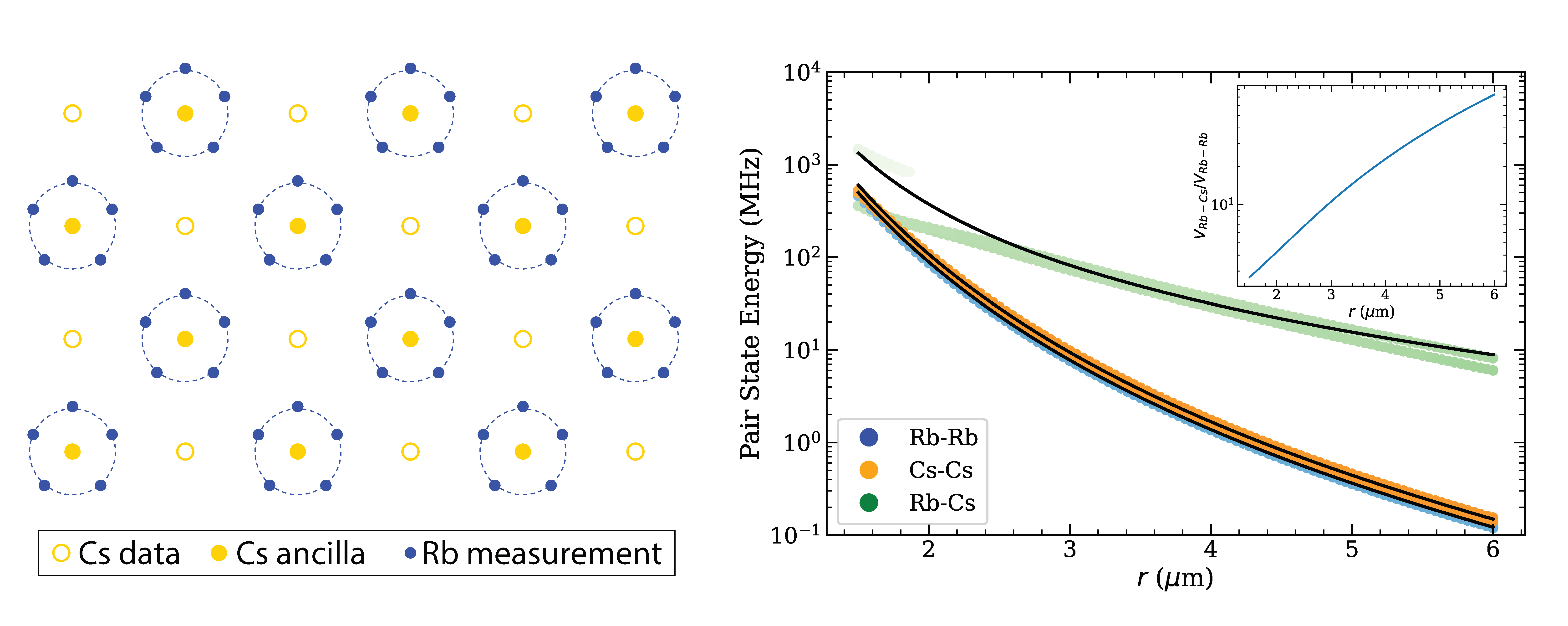}
\caption{\label{fig.asymmetric} Fast state measurements using a two-species repetition code. (left) Surface code layout with Cs data and ancilla qubits and
Rb measurement atoms. (right) Interaction potentials for the Rydberg pair states
 Rb$46s$–Cs$48s$. Colored points are the
relative energies of eigenstates of the diagonalized Hamiltonian
(shading is proportional to the eigenstate overlap  with the target
state). Solid lines are  fits used to extract $C_3$ and $C_6$ coefficients. Inset: Ratio of interspecies Rb-Cs interaction energy to
the intraspecies Rb-Rb interaction energy. Adapted from  \cite{Petrosyan2024}. }
\end{figure}

It is interesting to note that the two-species architecture also provides a  path to very fast state measurements without requiring optical cavities. The idea analyzed in  \cite{Petrosyan2024} is to use the second species as a repetition code to provide $N$ times the photon flux for a code of $N$ atoms. This implies close to $N$ times reduction in the integration time needed to make a measurement.  Naively we would like to perform the mapping 
$$ 
    |\psi\rangle \otimes |00...0\rangle_N \stackrel{?}{\rightarrow} |\psi\rangle\otimes|\psi...\psi\rangle_N
$$
where the state $|\psi\rangle$ is copied onto $N$ qubits. 
This is of course not possible because of the no-cloning theorem \cite{Wootters1982,Dieks1982}. However, we can use a repetition code. If $|\psi\rangle=c_0|0\rangle+c_1|1\rangle$ then an allowed mapping is  
\begin{equation}
\left(c_0|0\rangle+c_1|1\rangle\right)\otimes |00...0\rangle_N\rightarrow  c_0|0\rangle\otimes|00...0\rangle_N +c_1|1\rangle\otimes|11...1\rangle_N. 
\end{equation}
This mapping can be achieved by invoking asymmetric Rydberg interactions \cite{Saffman2009b}.  The ancilla qubit to be measured is encoded in one species, say Cs, and the auxiliary $N$ atoms for the repetition code are say Rb atoms, as shown in Fig. \ref{fig.asymmetric}.
Choosing Rydberg states that provide a  strong interspecies interaction and a weak intraspecies interaction the mapping can be obtained by exciting state $|1\rangle$ of the ancilla to a Rydberg state that blocks the state transfer of the auxiliary atoms via interspecies blockade.  Calculations show that with realistic experimental parameters preparation of the repetition code state can be done in under $5~\mu\rm s$ for $N\sim 5$ atoms, which should not be taken as a limit on the possible value of $N$. A related idea using electromagnetically induced transparency with a singe species was demonstrated in  \cite{WXu2021}.

Neutral atom mid-circuit measurements have only been demonstrated relatively recently and are still being actively developed. As such it is too early to say which method will work best at scale, and which atomic element will provide the best performance in terms of highest measurement fidelity, lowest loss, and shortest measurement time. Measurement times for arrays imaged in free space are still orders of magnitude longer than gate times, as can be seen in Fig. \ref{fig.measurement}. Fast cavity assisted measurements have been demonstrated in a few $\mu\rm s$ \cite{Bochmann2010, Grinkemeyer2025}, but integrating large atom arrays with cavities remains an outstanding challenge.

\section{Outlook}
\label{sec.outlook}

As neutral atom qubit arrays have matured attention has shifted to challenges that were not the center of attention early on. Individual atoms are natures ideal qubits; identical, well understood, and possessing of states that can be precisely controlled and which provide exquisite coherence. Nevertheless there are aspects of atomic qubits which open new error channels that don't exist in superconducting circuits or semiconductor quantum dots. 

The first is motional heating which   occurs due to photon recoils during gate operations and cannot be avoided if atoms are transported too rapidly. As seen in Fig. \ref{fig.transport_gate} the amount of heating per operation can be very small, but when extrapolated to a deep circuit with thousands or even millions of gates it is clear that recooling will be required.  The second is leakage into additional atomic states outside the qubit encoding basis. Leakage does occur in other platforms such as superconductors but it is particularly prevalent in atoms that have a plethora of accessible states. Even the $^{171}$Yb atom which has a spin $1/2$ nucleus and a spin singlet ground state providing an ideal two-level system can get trapped in metastable triplet and Rydberg states. Third there is atom loss due either to ejection from the optical trap due to a background gas collision, photoionization from a Rydberg state, or simply motion away from the trapping location when placed in an untrapped state during a gate operation. Together heating, leakage, and loss (HLL) represent a set of challenges that will need to be solved to realize the full potential of neutral atoms. 

In parallel with the progress on qubit count scaling, measurement fidelity, and gate fidelity there has been substantial recent work on HLL challenges. The heating challenge is common to both trapped ion and neutral atom qubits. In the trapped ion platform it is possible to use sympathetic cooling between two species \cite{Barrett2003} without affecting the quantum state encoded in internal degrees of freedom. A related idea has been proposed for alkaline earth like and alkali atoms based on state insensitive Rydberg interactions \cite{Belyansky2019}, as well as direct laser cooling in a regime that preserves the nuclear spin state \cite{Reichenbach2007}. Another, and  simpler approach is to invoke a two-species architecture as discussed in Sec. \ref{sec.midcircuit}. Provided the two species couple to very different optical wavelengths  they can be independently cooled, while preserving quantum coherence in the other species as has been shown
in  \cite{Singh2023}. Two-species maps naturally onto a logical qubit architecture with one-species acting as data and the other as ancillae. Swapping quantum information back and forth between the species, the other element can be periodically recooled which will allow for operation of deep circuits. 

Leakage outside of the computational basis occurs primarily due to polarization errors on gate control beams, and  decay from electronically excited and Rydberg states during gate operations.   There are approaches for repumping leakage into a depolarization error \cite{Carr2013,ICong2022}. To the extent that leakage can be detected or converted into loss at a known array location the leakage becomes an erasure error which can be used to significantly improve the performance of error correcting codes \cite{YWu2022,SMa2023, Scholl2023,MChow2024}. In the limit where a large fraction of all errors are erasures, and the erasures are biased which means they arise from leakage from only one of the computational qubit states,  the code threshold can be increased by close to an order of magnitude \cite{Sahay2023}. 

Lastly there is the problem of atom loss. One source of atom loss is collisions with untrapped background molecules. This loss mechanism can be reduced using cryogenic pumping to achieve  lower pressure and single atom lifetimes reaching 6000 s have been demonstrated \cite{Schymik2021, 
 Pichard2024,ZZhang2025}. With good vacuum and negligible background loss there will still be loss due to atoms being left in Rydberg states during gate operations. Loss may be partially mitigated by design of codes that can tolerate missing qubits \cite{Vala2005b,Baker2021, Perrin2025,ZWei2025}. Nevertheless a scalable system capable of running long calculations will need to incorporate a mechanism into the control system for monitoring and replacing lost atoms. First demonstrations of this capability that are able to maintain 1000 qubit scale arrays continuously occupied have been demonstrated \cite{Gyger2024, Norcia2024, NCChiu2025}. Fully integrating loss detection and atom replacement into the error correction cycle remains an outstanding challenge.

This contribution has strived to cover many, if not all, of the most active areas of development of the neutral atom qubit approach to quantum computing.
Figure \ref{fig.hpnisq} suggests 
two scenarios for the next phase of progress beyond NISQ towards broadly useful quantum accelerated computing. The HP-NISQ path requires substantial progress in gate fidelity. The FTQC  path requires substantial progress in scaling of the system size and control, and less so in raw fidelity. The HLL challenges discussed above will need to be solved in order to realize deep FTQC circuits. Alternatively there may be a path to solving some useful problems with HP-NISQ, without fully retiring HLL errors, and without the large overhead of the full machinery of quantum error correction. From that perspective it is possible that HP-NISQ machines which are useful for some specific applications will be available much earlier than FTQC. 

Taking a step back and dispassionately evaluating the field one is impressed by the wealth of new ideas and results that seem to emerge at an ever increasing rate. Admittedly these pages have devoted more space to specifying limitations and analyzing challenges than to applauding the progress. That should not be taken as an indication of pessimism, but as a recognition that it is by taking the challenges seriously that new solutions will be found. Lay people like to ask when we will have a really useful quantum computer. It is a hard question to answer, particularly for the experts who know all the challenges. The rapid progress with neutral atom qubits has attracted many new researchers to the field in the last few years, and that is a convincing  sign that we are on a path to achieving something important.

\vspace{.5cm} 

I would like to thank my research group at UWM and colleagues at Infleqtion for many stimulating discussions, Matt Otten for comments on the manuscript, and Sam Norrell for technical assistance.  
The work in Madison on neutral atom quantum computing is generously supported by the US National Science Foundation, the Department of Energy, the Laboratory for Physical Sciences, the Army Research Office, the Wisconsin Alumni Research Foundation, and Infleqtion.

%
%
\bibliographystyle{spphys.bst}
\bibliography{qc_refs,saffman_refs,rydberg,atomic,optics}

\end{document}